\documentclass[%
reprint,
superscriptaddress,
nofootinbib,
amsmath,amssymb,
aps,
pra,
floatfix,
10pt
]{revtex4-2} 
\usepackage{amsmath,amssymb}
\usepackage{mathrsfs} 
\usepackage[capitalise]{cleveref}
\usepackage{siunitx}
\usepackage{braket}
\usepackage{calc}
\usepackage{tabularx}
\usepackage{dsfont}
\usepackage{color}
\usepackage{seqsplit}
\usepackage{ifthen}
\usepackage{graphicx}
\usepackage{mathtools} 

\allowdisplaybreaks

\newcommand{\dif}{\mathrm{d}}%
\newcommand{\Nabla}{\vec{\nabla}}%
\newcommand{\ZT}[1]{\textquotedblleft#1\textquotedblright}%
\newcolumntype{Y}{>{\centering\arraybackslash}X}%
\newcolumntype{Z}{>{\raggedright\arraybackslash}X}%

\newlength{\myl}%
\newcommand{\SUM}[2]{{\setlength{\myl}{\widthof{$\displaystyle\sum_{#1}^{#2}$}*\real{0.5}-\widthof{$\displaystyle\sum$}*\real{0.5}}\sum_{#1}^{#2}\;\hspace{-\the\myl}}}
\newcommand{\INT}[3]{\settowidth{\myl}{$\displaystyle\int_{#1}^{#2}$}{\int_{#1}^{#2}\;\;\;\hspace{-\the\myl}\dif #3}\,}
\newcommand{\TINT}[3]{\settowidth{\myl}{$\int_{#1}^{#2}$}{\int_{#1}^{#2}\!\ifthenelse{\equal{#1#2}{}}{}{\;\;\;\;\hspace{-\the\myl}}\dif #3}\,}%
\newcommand{\EINT}[3]{\settowidth{\myl}{$\int_{#1}^{#2}$}{\int_{#1}^{#2}\;\;\;\,\hspace{-\the\myl}\dif #3}\,}

\begin{document}
\title{A review of active matter reviews}

\author{Michael te Vrugt}
\affiliation{DAMTP, Centre for Mathematical Sciences, University of Cambridge, Cambridge CB3 0WA, United Kingdom}

\author{Raphael Wittkowski}
\email[Corresponding author: ]{raphael.wittkowski@uni-muenster.de}
\affiliation{Institut f\"ur Theoretische Physik, Center for Soft Nanoscience, Universit\"at M\"unster, 48149 M\"unster, Germany}

\begin{abstract}
In the past years, the amount of research on active matter has grown extremely rapidly, a fact that is reflected in particular by the existence of more than 600 review articles on this topic. Moreover, the field has become very diverse, ranging from theoretical studies of the statistical mechanics of active particles to applied work on medical applications of microrobots and from biological systems to artificial swimmers. This makes it very difficult to get an overview over the field as a whole. Here, we provide such an overview in the form of a metareview article that surveys the existing review articles on active matter. Thereby, this article provides an introduction to the various subdisciplines of active matter science and constitutes a useful starting point for finding literature about a specific topic. 
\end{abstract}
\maketitle

\section{Introduction}
\begin{figure}[thbp]
\centering\includegraphics[width=\linewidth]{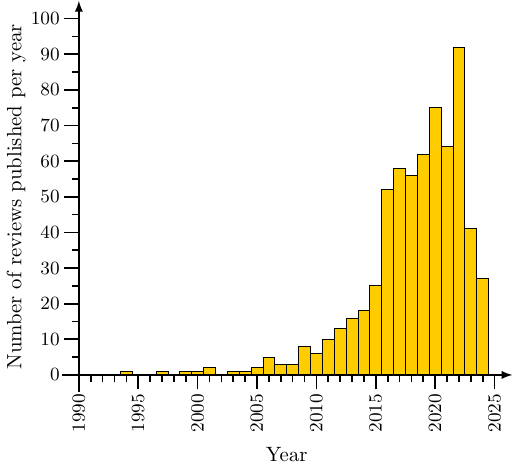}
\caption{\label{fig1}Number of reviews on active matter published per year. Very recently published reviews may not always be incluced.}
\end{figure}

The study of active matter (a broad class of locally driven nonequilibrium systems for which self-propelled particles are the most prominent example \cite{teVrugtLC2024}) is one of the most rapidly growing fields of physics, chemistry, and engineering. This is reflected not only by the amount of research articles published on this topic, but in particular also by the existence of several hundred review articles. See Fig.\ \ref{fig1} for a histogram showing the increasing number of reviews on active matter published per year. The number of publications on and subtopics of active matter science has reached an extent that makes it very difficult to get an overview over the entire field, which ranges from theoretical work on the collective dynamics of bacteria and artificial active colloids to applied studies aiming at the development of nanorobots for targeted drug delivery.

A way to get such an overview is a \ZT{review on reviews} (metareview) about active matter. This format is well established in medicine due to its usefulness for clinical practitioners who are otherwise not able to quickly find relevant literature. However, to the best of our knowledge, no such metareview has been written on active matter (or on any other topic in physics) so far. Since each review covers a large portion of the existing literature on active matter, a metareview allows to give a much broader overview over the work on active matter than standard reviews. For example, a researcher who is working on theoretical aspects of active particles, but is interested also in medical applications, could take a metareview as a starting point for finding literature about this aspect.

\begin{figure*}[thbp]
\centering\includegraphics[width=\linewidth]{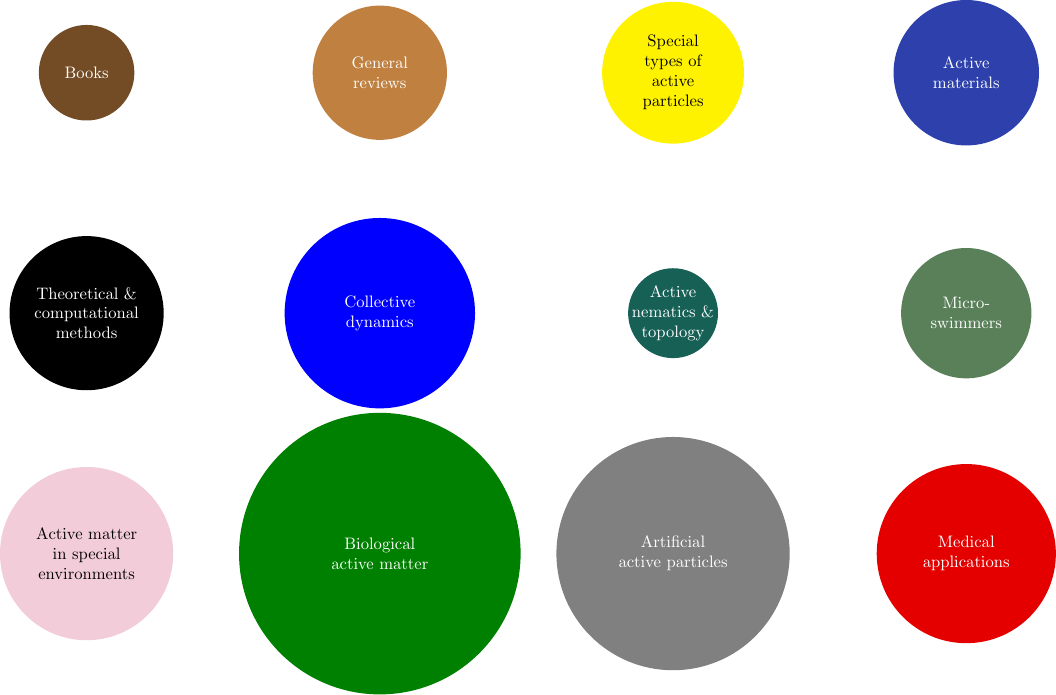}
\caption{\label{fig2}Published reviews on active matter grouped according to their types and the addressed subfields of active matter science. The areas of the disks are proportional to the number of reviews. Each disk corresponds to one of the sections \ref{sec:books}-\ref{sec:medical}. We do not show a circle for section \ref{colloids} (active colloids), as this term appears in many reviews across different types and subfields, and for section \ref{sec:molecular} (molecular active matter), as we have not completely surveyed this field. Reviews are not counted twice in this figure.}
\end{figure*}

This article constitutes such a metareview. It is divided into several subsections corresponding to different types of reviews and subfields of active matter science. Each subsection provides both a brief introduction to this subfield and an overview over the existing (review) literature. See Fig.\ \ref{fig2} for the numbers of reviews corresponding to these subsections. 

The metareview format has the following advantages:
\begin{itemize}
\item Given that there are many thousands of articles on active matter, it is essentially impossible for newcomers to get an overview over the entire field. Even a general review article has to be extremely selective in the literature it presents. A metareview presents an excellent starting point for finding literature on a specific subtopic one is interested in.
\item Review articles are only written on topics on which there is a lot of work and which editors and authors judge to be relevant and timely. Thereby, identifying topics that are reviewed a lot helps to find trends and promising topics for future work in a way that is not strongly affected by the personal judgement of individual review article authors.
\item For editors and review authors, a metareview allows to check whether or not a future proposal for a review fills an existing gap, and which topics are still worth being reviewed.
\item Due to its size and diversity, the field \ZT{active matter} has by now split into several disjunct subcommunities that are working relatively separately and use different terminologies (such as \ZT{active particles} vs \ZT{nanorobots}).  Given that a metareview allows to cover a larger topical breadth than other forms of review, it can help to create connections between different subgroups.
\end{itemize}

All articles cited in this review are review articles on active matter. This allows us to avoid having to introduce every article as \ZT{a review of motility-induced phase separation can be found in Ref.\ \cite{CatesT2015}}. Instead, a comment of the form \ZT{motility-induced phase separation is interesting \cite{CatesT2015}} should be understood as indicating that Ref.\ \cite{CatesT2015} is a review on motility-induced phase separation. Note that it is of course not clearly defined what an \ZT{active matter review} is -- some of the reviews cited here might be counted more as biology reviews, some of the articles might be considered to be not a review in the narrow sense, but something similar (a perspective), and some of the articles discuss active matter prominently but have a different overall topic (such as a certain theoretical method that is used not only in active matter science). We have understood the term \ZT{active matter review} in a broad sense for the purpose of writing this review in order to ensure that we give a broad overview. Moreover, we usually (although not always) discuss reviews only in one section to make the presentation more clearly arranged, even though some could be discussed in several ones (a review on collective dynamics in biological active matter, for instance, could in principle be discussed both in Section \ref{collectivedynamics} (collective dynamics) and in Section \ref{collectivedynamics} (biological active matter), but is here sorted in one of them).

\section{\label{sec:books}Books on active matter}
Books often aim at providing a general introduction to the basics of a field rather than over the latest developments within a certain subfield. When books are being published on a certain topic, this indicates that the corresponding field has reached a certain maturity, since a book takes a rather long time to write and some author needs to have the opinion that the foundations of a field are sufficiently developed and of interest to a sufficiently broad audience to be worth this effort. Books can be classified into monographs (all chapters have the same author or authors) and edited volumes (different chapters have different authors). 

Several books on active matter have been published by now, including some monographs \cite{Toner2024,SaguesMestre2024,Pismen2021,Schweitzer2003,Lauga2020} covering colloidal and biological active matter and collective dynamics as well as one monograph \cite{Berg2008} on the movement of Escherichia coli. Reference \cite{WittkowskitV2024} is an elementary physics textbook briefly discussing active particles. Monographs have also been written on the development of microrobots \cite{Bellouard2009,Sitti2017}. Other books on active matter are edited volumes. References \cite{TailleurGMYS2022,KurzthalerGS2023} contain reviews providing a broad overview over the field. A series of three books edited by \citet{BellomoDT2017,BellomoDT2019,BellomoCT2022} discusses theoretical modeling approaches with a focus on collective behavior. References \cite{Tibbits2017,teVrugt2024} are edited books discussing active matter as a potential route to the realization of programmable and intelligent materials. Finally, Ref.\ \cite{FratzlFKS2021}, which also approaches the topic from a materials science perspective, takes a slightly unconventional approach by focusing also on interviews and (philosophical/historical) essays.

\section{\label{generalreviews}General active matter reviews}
The next important class of literature is that of general active matter reviews, i.e., articles that aim to cover the entire field or that at least have no focus on a specific subtopic. Notable here are two reviews written by \citet{MarchettiJRLPRS2013} and \citet{BechingerdLLRVV2016} that have appeared in \textit{Reviews of Modern Physics} and can, due to their scope and the large number of citations (more than 3900 and 2500, respectively, according to Google Scholar on 19 May 2024), rightfully be considered standard texts in this field. Additional broad reviews written before 2020 include Refs.\ \cite{EbelingS2005,Menon2010,Menzel2015,YadavDBS2015,Nagel2017,Ramaswamy2017,Ramaswamy2019,DauchotL2019,Klotsa2019}. Reviews of this type continue to be written today \cite{DasSM2020,JiangH2022,BowickFMR2022,EssafriGDD2022,GentileKS2023,teVrugtLC2024}. Another type of general (in the sense of \ZT{not focused on a specific subtopic}) review is articles focusing on the latest trends at the time of their publication
\cite{PanH2017,GompperEtAl2020,AgostinelliCDDHMNSY2020}. Some general active matter reviews aim at nonspecialist audience \cite{Stark2007,Popkin2016,WilczekHB2021,LiebchenL2022,Sinha2024}. Finally, also editorials \cite{CatesM2011,Pismen2014,GompperBSW2021,Das2021} can -- at least in a broader sense -- be classified as general review articles. 

\section{\label{colloids}Active colloids}%
One of the most important types of active matter is \textit{active colloids}. A colloid consists of particles with a size much larger than atoms or molecules that are dispersed in a liquid or gas \cite{Aranson2013}. (Sometimes, the word \ZT{colloid} is also used to refer to colloidal particles rather than to the suspension as a whole.) An active colloid is a colloid where the dispersed particles are active. Owing to the importance of active colloids, a number of active matter reviews focus specifically on them \cite{Aranson2013,Ebbens2016,ZhangLGG2017,Poon2013,BishopBB2023,WangLMDZ2020}.

\section{\label{specialtypes}Special types of active particles}
A number of reviews focus on the dynamics of specific types of active particles. The most important of these, such as colloidal particles, microswimmers, and microrobots, are covered in other sections of this review. Here, we discuss reviews focusing on less common variants.

Among these, motile droplets are notable. Here, the active particle is a droplet of liquid immersed in another liquid. From a chemical perspective, such systems can be thought of as a special type of active colloid (see Section \ref{colloids}), namely an active emulsion \cite{BirrerCZ2022,IgnesJS2020}. An emulsion is a colloid where both the dispersed particles and the surrounding medium are liquids (the dispersed particles then take the form of droplets). Spontaneous motion of the droplets can be caused by a variety of mechanisms \cite{MaassKHB2016}. Typically, these droplets are covered by surfactants. The interaction of these surfactants with the droplet and the surrounding fluid gives rise to surface tension gradients and thereby to Marangoni flows, as a consequence of which the droplet moves \cite{MaassKHB2016,Michelin2022}. This propulsion requires a broken spatial symmetry, resulting (given the spherical symmetry of the particles) from spontaneous symmetry breaking in a bifurcation \cite{DwivediPM2022}. Reviews on active droplets either take a general perspective \cite{MaassKHB2016,SeemannFM2016,LohseZ2020,BirrerCZ2022,DwivediPM2022,CarenzaGN2023} or focus on specific aspects such as chemically propelled droplets \cite{ToyotaSHIK2020,Michelin2022,DonauB2023}, collective dynamics \cite{TanakaNK2017}, droplets in lipid systems \cite{BabuKLPR2022}, interfacial tension \cite{BanKSN2017}, and technical applications  \cite{VcejkovaBHS2017}.

Moreover, phoretic active particles \cite{MoranP2017,AubretRP2017,FukuyamaM2020,Golestanian2023,PopescuUD2016,WangDAMS2013} are notable here. Phoretic transport generally refers to the motion of (colloidal) particles that is driven by a gradient in a scalar field \cite{MoranP2017}, which can, for example, be an electrostatic potential (electrophoresis), a concentration field (diffusiophoresis), or temperature field (thermophoresis) \cite{Golestanian2023}. Phoresis can be exploited to generate self-propulsion if the particle can itself generate the gradient it is driven by. This can take a variety of forms such as self-electrophoresis, self-diffusiophoresis, and self-thermophoresis \cite{MoranP2017}. The reviews in Refs.\ \cite{MoranP2017,Golestanian2023} provide a broad discussion of this field. Other reviews focus on more specific aspects such as patchy phoretic particles \cite{AubretRP2017}, phoretic particles at interfaces \cite{MalgarettiH2021}, self-diffusiophoresis \cite{PopescuUD2016,FukuyamaM2020}, self-thermophoresis \cite{KroyCC2016}, and thermophoresis in Janus particles \cite{BresmeOCAG2022}.

Further types of \ZT{special} active particles reviewed in the literature are
\begin{itemize}
\item active polymers, which include enzymes, biological filaments, chains of Janus particles, and passive polymers in active environments \cite{WinklerEG2017,GoswamiCC2022,ZhuGSWY2024}.
\item active matter with inertia, which includes many macroscopic active particles (cars, flying animals, robots,...), activated dusty plasmas, and vibrated granulars \cite{Loewen2020}.
\item particles with anisotropic shape \cite{WensinkLMHWZKM2013,NagaiHT2015,WittenD2020,WangMLWLWW2022}. This includes in particular active rods \cite{BaerGHP2020,Peruani2016}, the perhaps most simple and important class of anisotropic active particles, and deformable active particles, which include active droplets and moving cells \cite{BoltzK2016,Ohta2017,Tarama2017}.
\item social systems \cite{CastellanoFL2009} such as crowds \cite{RadiantiGBSY2013,AlbiBFHKPPS2019,GongHPV2022,CorbettaT2023}, cities \cite{GallottiSD2021}, and traffic \cite{AlbiBFHKPPS2019}.
\end{itemize}

\section{Active materials}
Active particle systems constitute a new class of materials, and several review articles discuss them from a materials science perspective \cite{BernheimGST2018,Dufresne2023,dePabloEtAl2019,PattesonAJ2022}. A typical and important example is the review by \citet{FletcherG2009} on active biological materials, where the authors compare them to conventional materials. They argue that, while a good understanding of conventional materials can be gained by solely focusing on their bulk properties, active materials are governed by an interplay of processes occurring on many different time and length scales. A more recent materials science review by \citet{NeedlemanD2017} also emphasizes this aspect -- while self-organization in equilibrium requires small constituents (since the constituents can access different configurations only via thermal fluctuations), active systems form structures on all scales (molecules assemble in a cell, cells assemble in organs, organs assemble in an organism, organisms assemble in swarms).

Some reviews focus on specific types of active materials, such as \textit{active gels} \cite{ProstJJ2015,AlvaradoSSMK2017}. Gels are characterized by cross-linked systems, for which the cytoskeletal filaments of a cell are an example \cite{Liverpool2006,JoannyP2009}. The presence of molecular motors, however, makes the properties of such a gel different from that of a passive one \cite{ProstJJ2015}. Moreover, \textit{biological tissues} can also be studied using methods from materials science and thereby be considered as active materials \cite{XiSDLL2019,FrancoisB2023,EderSBF2021,NguyenSV2018,BurlaMVAK2019,HannezoH2022}. \textit{Active glasses} \cite{Janssen2019,BerthierK2023,BerthierFS2019} are another interesting type of active material. Generally, glasses are systems that behave like solids without possessing long-range order, often produced by supercooling a liquid to a point where the viscosity increases significantly due to particles being trapped \cite{Janssen2019}. Glass formation has been found in a number of biological and artificial active matter systems \cite{Janssen2019}. Reference \cite{MannaLSB2022} reviews the physics of \textit{active sheets}, two-dimensional active materials that can change their shape.

\textit{Liquid crystals}, an intermediate state of matter that has properties of both liquid and solid \cite{ZhaoEtAl2019}, are an interesting class of materials in the context of active matter \cite{BukusogluBMWA2016}. They consist of anisotropic (often rod-like) particles and can therefore form phases with orientational order. This phenomenon is also 
ubiquitous in living systems in the form of biological liquid crystals, which can be intracellular structures or assemblies of elongated cells \cite{ZhaoEtAl2019}. While passive liquid crystals are typically apolar (the rod-like particles do not have a preferred direction), active liquid crystals (liquid crystals consisting of active particles) can also be polar. An example for this is actin-myosin systems \cite{DogicSZ2014}. Active liquid crystals promise to have interesting applications due to their tunable properties \cite{ZhangMdP2021}. Also, active systems can be realized based on liquid-crystal-enabled electrokinetics \cite{PengL2019}. (See Section \ref{nemandtop} for a further discussion of liquid crystals.)

Finally, some reviews cover the development of \textit{intelligent matter} \cite{KasparRvdWWP2021,teVrugt2024} (materials possessing basic features of intelligence). In this context, nonequilibrium properties are essential \cite{KasparRvdWWP2021,Walther2020,MerindolW2017}, such that active matter is a particularly promising approach \cite{JeggleW2024}. Neural networks based on active agents are a special case of \ZT{liquid brains} (neural networks without a fixed spatial structure) \cite{PineroS2019}. Basic features of intelligence have also been found in slime moulds \cite{BoussardFODD2021,FreibergWK2024}. Biological active matter has been suggested to be able to perform self-organized computations \cite{England2022,teVrugt2024b}, and the same should be achievable with soft active materials \cite{Fischer2020} and robotic swarms \cite{OKeeffeB2019}. Consequently, an important aim of current work on microrobots is to make them intelligent \cite{SotoEtAl2021,MastrangeliMM2011,LoewenL2024}. Closely related is the concept of \textit{programmable (active) matter}, where external input allows to modify material characteristics \cite{YuFZCQSW2023,LozanoBB2023}.

\section{Theoretical and computational methods}
An important part of active matter science is the development of theoretical models for the description of active systems, and a number of reviews have been written specifically on this issue. While some articles do this on a rather general level \cite{ShaebaniWWGR2020,HechtUL2021}, most such reviews focus on a specific subtopic. References \cite{RomanczukBELSG2012,MarchettiFHPY2016}, for example, deal with \textit{active Brownian particles}, which is a popular example for a  \textit{particle-based model}. Here, one writes down equations of motion for every single active particle. 

\textit{Active field theories}, reviewed in Refs.\ \cite{TonerTR2005,JulicherGS2018,CatesT2018,Cates2019,teVrugtLW2020,ShaebaniWWGR2020,Loewen2021,MahdisoltaniG2021,CatesFMNT2022,Schmidt2022,teVrugtW2022,teVrugtBW2022}, are a very powerful theoretical method for describing the collective dynamics of active matter systems with many constituents. Here, one chooses a small set of order parameter fields (such as the density or velocity) and then describes the system using a partial differential equation for these order parameter fields rather than by solving the equations of motion for every single particle. Thereby, field theories provide easier access to analytical insights and qualitative understanding. Moreover, they allow to efficiently simulate systems with very large particle numbers. In addition to the general reviews in Refs.\ \cite{JulicherGS2018,Cates2019}, existing reviews cover theories for flocking \cite{TonerTR2005,Toner2023}, a field theory for self-chemotactic particles \cite{MahdisoltaniG2021}, entropy production \cite{CatesFMNT2022}, and the interaction-expansion method (a microscopic derivation technique for field theories) \cite{teVrugtBW2022}. An important field theory in the context of active matter is classical \textit{dynamical density functional theory} (DDFT), which describes the time evolution of the one-body density of a (passive or active) fluid. General reviews of DDFT can be found in Refs.\ \cite{teVrugtLW2020,teVrugtW2022}. Further reviews cover DDFT for microswimmers \cite{Loewen2021} and extensions of DDFT that do not require close-to-equilibrium assumptions (power functional theory) \cite{Schmidt2022,delasHerasZSHS2023} or provide a general overview over nonequilibrium coarse-graining techniques \cite{Schilling2022}.

While the phase behavior of passive systems can be described using the principles of equilibrium \textit{thermodynamics} -- the system is in or approaches the minimum of a free energy functional, and its phase behavior can be understood based on studying this functional -- the same is not (in general) true for active systems. Due to the usefulness of thermodynamic approaches for understanding equilibrium systems, much effort has been directed at understanding to which extent thermodynamic principles can be extended to active systems \cite{TakatoriB2016}. Reviews on this topic have covered, e.g., heat conduction \cite{ShomaliKVKG2022}, the notion of temperature \cite{PuglisiSV2017}, and synthetic nanomotors \cite{GaspardK2019}. \textit{Stochastic thermodynamics}, which allows to apply concepts from thermodynamics to small and noisy systems, has recently become an important tool in the field \cite{Seifert2012,Seifert2019,Maes2020,GnesottoMGB2018}. It can be used, e.g., for the study of entropy production \cite{CatesFMNT2022,Frenkel2023} and thermodynamic irreversibility \cite{OByrneKTvW2021,FodorLC2021}, but also in the study of nonequilibrium materials \cite{NguyenQV2021}. 

There have also been a few reviews that specifically focus on \textit{computational and numerical methods} for active systems. An overview over this topic is provided in Refs.\ \cite{ShaebaniWWGR2020,SabassEtAl2023}. Machine learning is a topic that has attracted particular interest in recent years, also in the context of active matter \cite{CichosGMV2020,Volpe2024,ZhangGJ2024,BrucknerB2024}. Other reviews have covered Lattice Boltzmann methods \cite{CarenzaGLNT2019}, microfluidic simulations \cite{KarampelasG2022}, particle swarm optimization \cite{Rezaee2014}, and simulations of active Brownian particles \cite{CallegariV2019}.

Further topics covered in modeling reviews are active colloids \cite{MaYN2020,ZoettlS2022,Yoshinaga2017,Hao2024,HuangWCT2024}, active suspensions \cite{SaintillanS2013}, and kinetic theory \cite{BellomoBD2009,Bianca2012}.

\section{\label{collectivedynamics}Collective dynamics}
A major reason for the interest of researchers in active particles in recent years is their collective dynamics. Systems of many interacting active particles can show \textit{emergent behavior}, i.e., behavior that qualitatively goes beyond what a single active particle would do \cite{ZoettlS2016,HaganB2016,StrogatzWYTADAG2022}. An early and very influential (more than 2100 citations according to Google Scholar on 19 May 2024) review on collective dynamics of active matter was published by \citet{Ramaswamy2010}.

The collective dynamics of active particles can differ in interesting ways from that of passive systems. \textit{Motility-induced phase separation} (MIPS) is a good example. In MIPS, particles with purely repulsive interactions spontaneously separate in a high-density and a low-density phase, something that would not be possible in a passive system. The reason for this effect is that the particles have a lower swimming speed in regions with high density, implying that they tend to accumulate there. MIPS was reviewed by \citet{CatesT2015} in an article that is today the standard reference for this topic. In addition, there exist a detailed review by \citet{BialkeSL2015} and several more recent introductory presentations \cite{OByrneTZ2021,Stenhammar2021,teVrugtBW2022}. The topic is also covered in a number of review articles that discuss the collective dynamics of active particles on a more general level \cite{Marenduzzo2016,Speck2016,Speck2020,Das2022,BinderV2021}.

Another important type of collective behavior is \textit{flocking}, which refers to the collective coherent motion of active particles \cite{TonerTR2005}. A paradigmatic example is a flock of birds, which can be modeled using techniques from theoretical condensed matter physics \cite{CavagnaG2014}. These techniques include particle-based approaches such as the Vicsek model \cite{Ginelli2016} and field theories such as the Toner-Tu model \cite{TonerTR2005}. Flocking can be understood in analogy to ferromagnetism, with the flight direction of the birds corresponding to the spin \cite{TonerTR2005}. It is reviewed in Refs.\ \cite{VicsekZ2012,WiandtKS2015,TonerTR2005,CavagnaGG2018,Ginelli2016,CavagnaG2014,Toner2023,AldanaLV2009}.

The study of analogies between active and quantum matter is also a growing field of research. Some reviews discuss effects arising both in active matter and in quantum systems (and then cover both of them). This includes non-Hermitian physics \cite{AshidaGU2020,LinlinEtAl2024eng}, odd viscosities \cite{FruchartSV2022}, and the Casimir effect \cite{DantchevD2023}.

Several reviews focus on \textit{oscillations and synchronization} in active matter. Hydrodynamic synchronization, reviewed by \citet{GolestanianYU2011}, is a paradigmatic example here. Here, active agents that undergo independent cyclic motion, such as cilia and flagella \cite{GilpinBP2020}, synchronize as a consequence of hydrodynamic interactions. This process is important for pumping and bacteria swimming. Further reviews on this topic are Refs.\ \cite{Friedrich2016,BruotC2016,UchidaGB2017,LaugaG2012}. Oscillations also play a role in other contexts in active matter science, such as in embryo development \cite{KruseR2011}, learning in slime moulds \cite{BoussardFODD2021}, and swarmalators (swarming oscillators) \cite{OKeeffeB2019}.

The term \textit{dry active matter} refers specifically to active systems where one does not explicitly take into account a fluid surrounding the active particles that they exchange momentum with. The opposite case would be \textit{wet active matter}. Dry active matter, in particular its collective dynamics, is discussed in a number of review articles \cite{Toner2023,Chate2020,ChateM2023,BenDorKT2023}.

\textit{Active fluids}, defined in Ref.\ \cite{Saintillan2018} as viscous suspensions of active particles, are also an active field of research. Here, active particle systems are understood as fluids and studied via methods from fluid mechanics. A fluid-dynamical phenomenon that has attracted much attention in recent years is \textit{active turbulence}. While turbulence in passive fluids occurs only when the Reynolds number (a dimensionless number characterizing flow properties) is very high, active systems (such as active nematics, see Section \ref{nemandtop}) can exhibit turbulence even at extremely small Reynolds numbers \cite{AlertCJ2021,ThampiY2016,SanjayJ2022,AlexakisB2018,Lowen2016}. Several reviews also treat the \textit{rheology} of active fluids \cite{Saintillan2018,LanzaroG2023,Fielding2023,PuertasV2014}. Active fluids can exhibit remarkable rheological properties, such as spontaneous flows or the existence of a superfluid state \cite{Saintillan2018}. Further topics covered in reviews are biological fluids \cite{BeppuIMS2018,KochS2011,Bees2020} and active flows on curved manifolds \cite{AlIzziM2021}.

\textit{Chemotaxis}, where particles follow a chemical concentration gradient, can give rise to interesting forms of collective dynamics and is therefore the topic of several reviews that discuss (among other things) collective behavior \cite{Stark2018,Camley2018,JiWPZ2023,LiebchenL2018,LiebchenL2019}. For example, chemotactic particles can form clusters via the Keller-Segel instability. Here, particles are attracted by chemical species that they themselves produce, leading to a positive feedback and thus to a collapse \cite{LiebchenL2018}.

The \textit{collective dynamics of active colloids} is the topic of a number of reviews. Some address this issue on a rather general level, including also discussions of nonequilibrium colloidal systems in general \cite{Aranson2013b,Klapp2016,RoyallCDRSSV2024}. Other reviews focus on the \textit{(self-)assembly} of active colloidal systems, which can be tuned by changing properties such as their shape \cite{MalloryVC2018,LiZH2016,SotoWDD2021,ArangoB2019,WangDASM2015}. The collective behavior of active colloids depends on their \textit{interactions}, which can take interesting forms such as novel hydrodynamic or nonreciprocal interactions \cite{LiebchenM2021,Klapp2023,WinklerG2020}.

Further reviews on collective dynamics of active matter concern active chemical reactions \cite{Zwicker2022}, chiral active matter \cite{LiebchenL2022b}, frustrated active matter \cite{OrtizAmbrizNRRT2019}, hydrodynamically induced collective motion \cite{Kimura2017}, jamming \cite{ReichhardtR2014}, model experiments \cite{Dauchot2023}, multi-robot systems \cite{OhSSJ2017,GoldmanZ2024}, and planar active matter \cite{CugliandoloG2023}.

\section{\label{nemandtop}Active nematics and topology}
The study of liquid crystals consisting of rod-like particles has a long tradition in soft matter physics. Such particles often form so-called \textit{nematic phases}, in which the particles align their orientations. Since the rods have no preferred direction, such systems do not have polar order. An important feature of such liquid crystals is the existence of \textit{topological defects}, which are points (lines, planes) where the nematic order parameter changes discontinuously.

These concepts have also been transferred to active matter science. An \textit{active nematic} consists of rod-shaped active particles. Important examples include rod-shaped bacteria and rod-like intracellular structures such as microtubule-kinesin mixtures \cite{DoostmohammadiIYS2018}. Typically, the particles in an active nematic are also apolar -- the activity does not consist in spontaneous self-propulsion on the single-particle level, but in self-generated stresses acting on the environment \cite{ShankarSBMV2020}. See Refs.\ \cite{DoostmohammadiIYS2018,Yeomans2023} for general reviews of active nematics and Ref.\ \cite{DierkingA2017} for a brief discussion of active liquid crystals.

Topological defects in active nematics differ in several ways from those in passive ones, in particular by spontaneous motion, creation, and annihilation \cite{DoostmohammadiIYS2018,ShankarSBMV2020}. Several reviews have focused specifically on topological defects in active matter \cite{Aranson2019,ShankarSBMV2020} and biological systems \cite{ArdavsevaD2022}. Specific aspects covered in such reviews include microbial systems \cite{Sengupta2020}, flows generated by topological defects \cite{BrezinRJ2022}, and solitonic and knotted active matter \cite{Smalyukh2020}.

Many reviews of active nematics have a biological focus since there are many biological examples for this type of active matter, in particular in cell biology \cite{DoostmohammadiL2021}. This includes, e.g., the cytoskeleton \cite{BalasubramaniamML2022}, microbial systems \cite{Sengupta2020}, and tissues \cite{BalasubramaniamML2022,SawXLL2018}. Other topics covered in reviews are active nematics in confinement \cite{SaguesGHMI2023}, lyotropic active liquid crystals \cite{DierkingM2020}, and oscillatory active nematics  \cite{MikhailovKK2017}.

\section{\label{microswimmers}Microswimmers}
An important field of active matter research is the study of \textit{microswimmers}. This term refers to small objects that move (swim) in a fluid. There are many biological examples for microswimmers (bacteria, algae, sperm,...) \cite{KrugerE2016,JeanneretCP2016}, but there also exist many artificial variants \cite{NiuP2018,Stark2021}. The size of a microswimmer leads to hydrodynamic properties that are very different from that of a macroscopic swimmer. This is a consequence of the fact that the Reynolds number for such particles is very small. At this scale, the fluid flow is governed by the time-independent Stokes equation, implying that the swimming motion has to be non-time-reversal-invariant (Scallop theorem) \cite{YeomansPS2014}. General reviews of microswimmer physics are given by Refs.\ \cite{LaugaP2009,Yeomans2017,YeomansPS2014,ElgetiWG2015}.

Several reviews also focus on specific examples of microswimmers. Sperm cells are a paradigmatic example of microswimmers. They are useful for studying chemotaxis of active particles, as they are guided by chemicals released by the egg \cite{GaffneyGSBK2011,KauppA2016,KholodnyyGCB2020}. The collective dynamics of sperms \cite{SchoellerHK2020} and their hydrodynamics in the context of reproduction \cite{FauciD2006} have also been studied. Further examples for microswimmers are Escherichia coli \cite{KumarP2010}, green algae \cite{Goldstein2015,JeanneretCP2016}, pelagic organisms \cite{Kiorboe2016}, planktonic microorganisms \cite{GuastoRS2012}, and trypanosomes \cite{KrugerE2016} (parasitic flagellates causing diseases). (Similar reviews exist also for \ZT{macroswimmers}, such as fish or whales \cite{FishL2006}.) Reference \cite{KurtzhalerS2023} discusses cell swimming in general. In addition to specific biological species, reviews can also consider microswimmers with special swimming properties such as chemotactic swimmers \cite{LiebchenL2019}, chiral swimmers \cite{Lowen2016,LiebchenL2022b}, and gyrotactic swimmers \cite{QiuMZ2022}. 

The behavior of microswimmers is also strongly influenced by their environment. A well-known example is the fact that microswimmers accumulate at surfaces \cite{ElgetiG2016}. Interfaces are, from a theoretical point of view, an example for an external field. Other external fields relevant for microswimmer physics are gravitational fields and harmonic traps \cite{Stark2016}.

Several reviews have also been published on simulations of microswimmers, covering fully resolved hydrodynamics \cite{OyamaMY2017} and multiparticle collision dynamics \cite{Winkler2016,Zottl2020}. Reference \cite{Ishimoto2023} discusses Jeffery's equations (equations used in microswimmer physics). Squirmers are covered in a recent review by \citet{Ishikawa2024}. On the experimental side, microfluidic techniques for microswimmers have been reviewed in Ref.\ \cite{SharanNLS2021}. A field that has been of particular interest in recent years is the construction of \ZT{smart} microswimmers that use reinforcement learning for navigation \cite{Stark2021,NasiriLL2023,LoewenL2024}.

\section{\label{specialenv}Active matter in special environments}
A notable portion of active matter research focuses on the behavior of active particles in special environments, a topic that is therefore also the subject of several review articles. Even Ref.\ \cite{BechingerdLLRVV2016}, which was listed in Section \ref{generalreviews} among the general reviews due to its importance for the field, focuses on \ZT{complex and crowded environments}. Active matter in crowded environments is moreover reviewed in Refs.\ \cite{TheeyancheriSKC2022,MartinezTD2023}. Reference \cite{GranekKKRST2024} provides a general discussion of the interaction of active systems with their environments.

An important type of \ZT{special environment} is \textit{confinement}, where the active particles are constrained by the geometry of the system they move in. Existing reviews on this aspect consider active nematics \cite{SaguesGHMI2023}, colloids \cite{KreuterSNLE2013}, microfluidics \cite{ChakrabortyMSD2022}, micromotors \cite{XiaoWW2018}, narrow channels \cite{AoGLSHM2014}, obstacle lattices \cite{ZhuWA2022}, and porous media \cite{KumarGA2022}. Moreover, active particles at \textit{surfaces} can show interesting dynamics that differs from their behavior in bulk, often they show accumulation  \cite{ElgetiG2016} (see also Section \ref{microswimmers}). Active matter at \textit{fluid interfaces} has been the subject of several reviews \cite{FernandezRodriguezRVCVHA2016,MalgarettiH2021,PandayBD2022,FeiGB2017,DengMCYRS2022,PopescuUDD2018}. Reference \cite{JinS2023} focuses specifically on microbes in porous environments, Ref.\ \cite{Snezhko2016} on torque-driven colloids at interfaces. 

Another example for a special environment an active particle can find itself in is an \textit{experimental setup}. Notable here are optical tweezers \cite{PolimenoEtAl2018} and outer space (where low-gravity conditions can be realized) \cite{VolpeBCGLSV2022}. The special environment can also be another interesting soft matter system, such as a complex fluid \cite{PattesonGA2016,IgnesS2022} or a liquid crystal  \cite{Lavrentovich2016,HernandezTIS2015}. Finally, the complexity of an environment can also be used to control the properties of active matter \cite{Aranson2018}.

\section{\label{bio}Biological active matter}
Biological systems are the paradigmatic example for active matter, and a significant portion of the research in active matter science is motivated by the aim to better understand biological systems. Consequently, a large number of reviews have focused specifically on living matter. Some of them simply provide general remarks on the general importance of the interface between physics and biology as a growing field of research \cite{Gardel2012,Needleman2015,RivelineK2017,Phillips2017,Bialek2017,Schwille2019}.

Biological active matter science is often concerned with cellular or subcellular systems \cite{Voth2017,TurlierB2019,Zidovska2020,LiTG2021,Nolte2024}. This starts at the very small (molecular) scales with work on the self-assembly of \textit{proteins} \cite{FreyB2023,Shelley2016}. Of particular importance here are \textit{enzymes}, catalytically acting proteins that can also be viewed as active matter \cite{FengG2020,GhoshSS2021}. Recently, there has also been some interest \cite{HeadBG2013,EltingSD2018,NazockdastR2020,AnjurKN2021,GouveiaSP2023} in the physics of the \textit{spindle apparatus}. The spindle apparatus is a cellular structure that is responsible for segregating the chromatids among the daughter cells during cell division \cite{NazockdastR2020}. Active matter science allows to understand mechanisms involved in its formation \cite{HeadBG2013,GouveiaSP2023}. \textit{Microtubules}, the main constituent of the spindle apparatus, have also been discussed in reviews \cite{BarsegovRD2017}. Further reviews at the interface of active matter and molecular biology have dealt with genetics \cite{ZhangW2017} and virus growth \cite{YinR2018}.

A central topic in this context is the physics of the \textit{cytoskeleton}, a biopolymer network that provides mechanical stability for the cell \cite{BauschK2006}. The cytoskeleton consists of actin, microtubules, intermediate filaments, and septins \cite{BashirzadehL2019}. From a physical point of view, the cytoskeleton can be viewed as an active gel \cite{Liverpool2006,JoannyP2009}, which is a viscoelastic material consisting of polar filaments that is in a nonequilibrium state \cite{JoannyP2009}. Thereby, the study of the cytoskeleton links cell biology with polymer and active matter science. Reviews of cytoskeleton physics can be found in Refs.\ \cite{JuelicherKPJ2007,BanerjeeGS2020,BauschK2006,Liverpool2006,JoannyP2009,KhanAAM2012,DeLaCruzG2015,AhmedFB2015,MurrellOLG2015,MogilnerM2018,FosterFSN2019,BashirzadehL2019,IerushalmiK2021,BinotSC2021,FurthauerS2022,LorenzK2022,ConboyPvdNK2024,IngberWS2014}.

A field of work that links cellular biophysics and collective animal behavior is the study of \textit{phase separation} in biology. In cell biology, liquid-liquid phase separation is particularly important. Cells are organized in compartments, which is required for separating the various biochemical processes happening inside them. Some compartments, such as nucleoli, do not have a membrane, which raises the question why they do not simply mix. The answer is that the compartments are formed by liquid-liquid phase separation \cite{HymanWJ2014,BentleyFD2019,SneadG2019,JulicherW2024}. Such effects are also of medical relevance \cite{WangEtAl2021,TongEtAl2022}. Closely related is \textit{biological pattern formation}, which occurs, e.g., in morphogenesis \cite{MaroudasYK2021,GrossKG2017,PfeiferSR2024} and in ecological pattern formation \cite{LiuRHPFvdK2016,Silverberg2016,HeffernHBI2021}, where different organisms distribute in space. Chiral structures are often observed in biological pattern formation \cite{RahmanPW2024}. A theoretical tool for modeling biological patterns is given by nonlocal diffusion equations \cite{PainterHP2023}. 

Another direction of research focuses more on \textit{(nonequilibrium) statistical mechanics in biology}, in particular on \textit{noise}. Biological systems are very noisy, and this noise plays a crucial role for their behavior  \cite{Tsimring2014,Ritort2019}. Moreover, biological systems are far from thermodynamic equilibrium, and this makes applying tools from nonequilibrium physics crucial for understanding them \cite{FangKLW2019}. For instance, phase transitions in biology can be fundamentally different from equilibrium phase transitions known from thermodynamics \cite{LeeW2018}. A related topic is the study of critical phenomena in the life sciences \cite{Munoz2018}. Information-theoretical concepts have also been considered relevant in this context  \cite{BasakSHHK2021,Funk2022,LanT2016,Walker2017}.

A central topic in biophysics is the physics of \textit{cellular motion}. An important type of cellular motion (besides swimming, see Section \ref{microswimmers}) is \textit{cell crawling}, the motion of cells on a surface. At the leading edge of the cell, rapidly growing protein chains (actin filaments) push the membrane outwards. The resulting extension then adheres to the surface. At the rear, the cell deadheres and contracts \cite{SchwarzS2013}. Reviews of cell motion and cell adhesion, which typically cover both biological aspects and physical models, can be found in Refs.\ \cite{SchwarzS2013,SackmannKH2010,CaballeroCPVR2015,MasuzzoVTAM2016,DengL2022,RafelskiT2004,Giomi2019,WickstroemN2018,Kumar2021}. An important mechanism in this context is chemotaxis \cite{LevineR2013}, where the cells navigate based on chemical gradients.

Going to a larger scale, the \textit{collective dynamics of cells} is an important subfield of active matter science \cite{VedulaRLL2013,Gov2014,HakimS2017,Brochard2019,GopinathanG2019,PajicM2021,LaPortaZ2019,PajicM2023,TonerTR2005,MehesV2013}. Collective cell migration is important, e.g., for understanding tissue formation, wound healing \cite{TurleyCLWM2022}, and tumor growth \cite{Rorth2009}. Due to the medical relevance of this topic, much work has focused specifically on cancer \cite{SpatareluEtAl2019,Uthamacumaran2020,BellomoD2008,BlauthKGGK2021}. Collective dynamics has, however, also been studied for other specific cell types such as epithelial cells \cite{ZornMSR2015}, stem cells \cite{JorgKYS2021}, and squamous cells \cite{GanjreKIP2017}. Another major topic in the work on collective cell migration is the development of theoretical models \cite{AlertT2020,ButtenschoenE2020,TliliGGMMS2015,YangJL2019,BanerjeeM2019,KosztinVF2012}. Cells can also divide and thereby exhibit nonconserved dynamics, as a consequence of which they are considered an example of \ZT{proliferating active matter} \cite{HallatschekDDDEWW2023}.

A crucial aspect in collective (and single) cell mechanics are \textit{forces}. These can be measured in a variety of ways \cite{GomezLAT2020}, and can in particular be obtained from cell deformations \cite{MerkelM2017}. Mechanical forces play a central role in morphogenesis \cite{Grill2011}, both in animals \cite{LegoffL2016,BraunK2018} and in plants \cite{MirabetDBH2011}. They can also be relevant for cell polarization \cite{LadouxMT2016,OkimuraI2017}. Reviews focusing specifically on the mechanobiology of collective cell behavior can be found in Refs.\ \cite{LawsonM2021,LadouxM2017,VishwakarmaSD2020,DumontP2014}.

A major part of biological active matter science is the study of \textit{bacteria} \cite{Aranson2022,JinS2023,Berg2000}. For physicists, the collective dynamics of bacteria is of particular interest \cite{FengH2017,Partridge2022,BeerA2019}. For instance, theoretical models for MIPS (see Section \ref{collectivedynamics}) are often motivated by simple descriptions of quorum-sensing bacteria \cite{CatesT2015,Cates2012} (i.e., bacteria who adapt their behavior to the population density \cite{MillerB2001}). \textit{Biofilms} -- accumulations of microorganisms and extracellular substances on a surface \cite{IshikawaOK2020} -- are a prominent example of collective phenomena in bacteria systems \cite{IshikawaOK2020,VaccariMNLLS2017,WongEtAl2021,GrobasBA2020}. Physicists are, however, not only interested in collective dynamics, but also in the biomechanics \cite{IshikawaOK2020,DufreneP2020} and hydrodynamics \cite{Lauga2016,WheelerSRS2019} of bacteria. Reviews have, in addition to addressing this topic, also covered specific types such as bacteria in the earth's surface \cite{JerolmackD2019}, big bacteria \cite{SchulzJ2001}, cyanobacteria \cite{VaruniMM2022}, magnetotactic bacteria \cite{KlumppF2016}, and meningococci \cite{BonazziEtAl2020}. A closely related topic is the physics of microswimmers, discussed in Section \ref{microswimmers}.

While the topics listed so far are concerned with rather small scales (cells or at most multicellular structures such as tissue), active matter science is also relevant on higher biological scales. An important field in this context is \textit{collective animal behavior}. The paradigmatic example, flocking of birds, has already been discussed in Section \ref{collectivedynamics}. But birds are not the only sort of animals that are studied in active matter science. \textit{Insects} are of interest here, not only due their material properties \cite{SchroederHWM2018}, but also due to their distinctive collective behavior \cite{FeinermanPGFG2018,Budrikis2021,ShishkovP2022,HuPAB2016} and flight mechanics \cite{Sun2023}. Moreover, \citet{LopezGCT2012} have reviewed the collective dynamics of \textit{fish}. Besides reviews focusing on such specific aspects, there have also been general reviews on collective behavior in biology \cite{SumpterMP2012,CavagnaGMW2018,Ouellette2022,DeC2022}.

\section{\label{sec:molecular}Molecular active matter}
Active matter science can be found on all scales -- from large ones (collective behavior of larger animals) to the very small molecular scales. The paradigmatic example for the latter are \textit{molecular machines}, for the design of which the Nobel prize 2016 was awarded. Consequently, a notable review on this topic is the Nobel lecture by \citet{Feringa2017}. Moreover worth mentioning are a review by \citet{KassemvLLWFL2017} on the design of molecular motors and a review by \citet{JulicherAP1997} on the development of theoretical models for them. 

Following the terminology from Ref.\ \cite{KassemvLLWFL2017}, a \textit{molecular machine} is a system in which (sub-)molecular components move as a consequence of a stimulus. A \textit{molecular motor} is a molecular machine in which the change of position of the components exerts an influence on a system. There are many biological examples for this, in particular motor proteins such as kinesin, dynein, and myosin. These move along cellular filaments in order to perform certain biological functions \cite{JulicherAP1997}. In addition, it is now possible to develop synthetic molecular motors \cite{Feringa2017,KassemvLLWFL2017}. Review articles on this topic cover both the biological  \cite{MorelandB2022,ZhaoGMS2018,MugnaiHJT2020,KolomeiskyF2007,LvYL2022,AgudoATIG2018,MerindolW2017} and the synthetic \cite{Feringa2017,KassemvLLWFL2017,MerindolW2017,ColbergRBK2014,Santiago2018,BrowneF2006,NaeemEtAl2021,GuoLWXL2020} case, as well as the \ZT{mixed} scenario of 
artificial systems that involve biological motors \cite{SaperH2019,Hess2011,KeyaKK2020,FeiL2022,ItoMNI2017,HessR2017,PatinoAMPS2018}.

\section{\label{arti}Artificial active particles}
A major field of research and the subject of a significant portion of the existing review literature is the development of artificial active particles. Review articles covering this topic on a general level are, e.g., Refs.\ \cite{FusiLLPvHA2023,KaturiMSS2016,LiuZ2021,EbbensH2010,KimGLZF2016,ParmarMKSSTPSS2015,AlarconWQF2016,YangXZYGHSY2020,FernandezRMHSS2020,Fujita1994,YuCWH2021,GuixWSMS2018,Santiago2020,XuGXZW2017,ShieldsWV2017,ChenEtAl2018}. A variety of approaches can be distinguished (which can also be combined in the same particle \cite{ChenSKLW2019,RenWM2018}).

\textit{Janus particles} are one of the most prominent types of artificial active particles. They are named after the Roman god Janus, who has two faces. Similarly, Janus particles have (at least) two different sides with different physical or chemical properties. This is important in the context of active matter since self-propulsion requires some kind of symmetry breaking that determines the direction in which the particles move. In the case of Janus particles, active motion can be achieved by coating with platinum on one side. When the particles are immersed in a hydrogen peroxide solution, the platinum catalyzes the decomposition of hydrogen peroxide. This leads to an anisotropic distribution of reaction products and thereby to self-diffusiophoresis driving the particles forward \cite{ZangGG2017}. Janus particles are reviewed in Refs.\ \cite{WaltherM2008,WaltherM2013,JuradoE2017,ZhangLG2015,ZangGG2017,CampuzanoGSPYP2019,Kurup2020,NizkayaAV2022,BresmeOCAG2022,LiuHXZ2022,DeyWAS2016,SanchezLJ2015}. 
 
A related topic is the \textit{light-driven propulsion} of active particles. This can be achieved by a variety of mechanisms \citep{SafdarSJ2017} such as light-induced surface effects, photo-thermal effects, momentum-transfer of photons, and light-induced chemical reactions. In many cases, phoretic propulsion (see Section \ref{specialtypes}) is relevant here. The light causes an anisotropic distribution of, e.g., electric flow, chemicals, or temperature, as a consequence of which the particle moves \cite{VillaP2019}. Such an asymmetry can be obtained by using Janus particles \citep{SafdarSJ2017}. For example, if a semiconducting particle is illuminated at sufficiently high photon energies, electrons are shifted to the conduction band, diffuse on the particle surface and then induce reactions with surrounding molecules. If the reaction products have an anisotropic distribution, it can give rise to self-electrophoresis or self-diffusiophoresis \cite{SipovaAJK2019}. General reviews of light-driven active particles are given by Refs.\ \cite{ReyVV2023,ZhanWZYLY2018,VillaP2019,SafdarSJ2017,SipovaAJK2019,PalagiSF2019,ZhouZCL2020,WangXT2021,VolpeEtAl2023}. Other reviews cover specific aspects such as BiVO$_4$ microparticles \cite{HeckelWRVS2024}, the control of collective dynamics \cite{ZhangGMG2018}, light-driven fluid micropumps \cite{ShklyaevMLKTSB2023}, molecular systems \cite{LanciaRK2019,Kageyama2019}, photocatalytic motors \cite{DongCYGR2018}, and photo-bioconvection \cite{JavadiATP2020}.

Moreover, active particles can be propelled via \textit{ultrasound}. Microparticles in a sound wave field in a fluid are subject to acoustic radiation forces. Primary radiation forces, arising from interactions between particle and sound, lead to particle movement, whereas secondary radiation forces, arising from waves reflected by the particles, lead to particle-particle interactions \cite{LiMOP2022}. General reviews of ultrasound-driven particles are given by Refs.\ \cite{RaoLMZCW2015,XuXZ2017,LuSZWPL2019,LiMOP2022,RenSHW2021,McneillM2023}. Among these, the review by \citet{McneillM2023} is the most recent one and gives a very general overview over both propulsion mechanisms and collective behavior. Topics covered in specific reviews include the combination of acoustic propulsion with chemical propulsion \cite{RenWM2018} and medical applications \cite{LealVSG2021}. In medical applications, particles propelled by ultrasound are particular promising since they do not require chemical fuels and can be controlled externally. A related topic is ultrasound imaging \cite{WangZ2020}.

\textit{Magnetic} micro- or nanoparticles are also worth discussing. If a magnetic particle with volume $V$ is in a magnetic field $\vec{B}$, it exhibits a magnetization $\vec{M}$ and is subject to a force $\vec{F} = V (\vec{M}\cdot\Nabla)\vec{B}$ and a torque $\vec{T} = V \vec{M}\times \vec{B}$ \cite{DongWIWN2022}. Thus, one can, for instance, propel a particle via magnetic field gradients \cite{HouWFWYZHSF2023}. This by itself is not active motion, but simply motion driven by an external field. Microrobots can be propelled forward via rotating fields if they have a helical structure -- the rotation around the helical axis is converted into a translational corkscrew motion \cite{ZhouMPZP2021}. Such approaches are based on the propulsion mechanisms of bacteria \cite{PeyerZN2013}. The application of oscillating magnetic fields allows to realize genuine active motion \cite{MandalPKG2018}. Magnetic actuation mechanisms are, due to the absence of chemical fuels, useful in medical contexts \cite{DongWIWN2022} (similar to the case of ultrasound-driven propulsion). Reviews of magnetically propelled micro- and nanorobots can be found in Refs.\ \cite{HouWFWYZHSF2023,JinZ2021,MandalPKG2018,ZhouMPZP2021,MeijerR2021,TiernoS2021,PeyerZN2013,KlumppLBF2019,DongWIWN2022} and cover several aspects such as biomedical applications \cite{PeyerZN2013,DongWIWN2022}, collective dynamics \cite{JinZ2021}, or preparation methods  \cite{MeijerR2021}.

It is also possible to propel active particles via \textit{electric fields}, a topic reviewed in Refs.\ \cite{BoymelgreenSKY2022,DiwakarKMYV2022}. \citet{BoymelgreenSKY2022} list four mechanisms of this form, namely (i) electric fields inducing a rotation in dielectric particles, (ii) particles breaking the symmetry of an electrohydrodynamic flow and thereby translating, (iii) induced-charge electroosmosis, and (iv) electrostatic self-diffusiophoresis.

A topic that deserves a specific discussion in the context of artificial active matter is the development of \textit{bioinspired systems} \cite{OhSSJ2017,ArenaBBFF2021,SantiagoS2019,BannoST2022,OhSSJ2017,NsamelaGMS2022,MallickKR2024}, where the particles are in some way supposed to mimic the behavior and working mechanisms of biological organisms. For instance, small self-propelled robots can use helical propulsion mechanisms inspired by those of bacteria \cite{PalagiF2018}. Bioinspired synthetic aggregates \cite{VernereyEtAl2019}, biomimetic sensors \cite{LiLZ2019}, biohybrid robotics (combination of biological systems with artificial materials) \cite{MestrePS2021,WangP2018}, and synthetic morphogenesis \cite{DaviesL2023} can also be classified in this category. A particularly notable subfield here is the development of artificial cells, a project that aims to contribute to our understanding of the functions of biological cells and of the origin of life \cite{JeongNKLS2020}. For example, work has been done here on the reconstitution of cell motility \cite{SitonB2016} (see Section \ref{bio} for a discussion of the biological version).

A closely related field is \textit{soft microrobotics} \cite{HuPN2018}. Soft robots are robots that can change their shape, which is useful in a variety of contexts such as motion control and adaptation to special environments \cite{MedinaMGFS2018}. This field has a close connection to bionics as many employed mechanisms are inspired by soft organisms like worms, jellyfish, and octopuses \cite{ZhangL2020}. For very small robots, adhesive forces and surface tension need to be taken into account \cite{WangHSZGPZ2022}.

Other topics covered in reviews on artificial active matter include 3d-printed microrobots \cite{DabbaghSMSST20223}, applications in chemistry \cite{DuanWDYMS2015}, artificial microswimmers \cite{NiuP2018,Degen2014,ParmarVS2016}, bioengineering applications \cite{CeylanGKS2017}, chemical pumps \cite{ShklyaevSB2018}, cargo manipulation \cite{WuBY2022}, colloidal motors \cite{ChenZW2019}, environmental applications \cite{GaoW2014,ZareiZ2018}, in vivo applications \cite{EstebanFernandezdeAvilaALGZW2018}, metal-organic frameworks \cite{TerzopoulouNCNPPL2020}, micro-/nanomotors \cite{LiuCLWZG2020,YuGZCZ2019,LinWWXH2016,Tu2017,WongDS2016,YuanLWM2021,ZhangSMGS2021,JohF2021,ZhangYZ2019}, nanozymes \cite{XieJFY2020}, and trajectory design \cite{AlHarraqBB2022,LinLWK2022}.

\section{\label{sec:medical}Medical applications}
A topic that in principle also falls under the topic \ZT{artificial active matter}, but that is discussed separately here due to its importance, is the study of \textit{medical applications}. A major inspiration for this subfield of active matter science in particular was Richard Feynman's idea that one could some day \ZT{swallow the surgeon} (the surgeon being a nanorobot that is able to perform medical interventions), an idea that also inspired the science-fiction movie \textit{Fantastic Voyage} \cite{Venugopalan2020,MundacaAFZW2023}.

In fact, based on the number of reviews covering this topic, it might be one of the largest subfields of active matter science -- although this fact is sometimes obscured by terminology, namely by the fact that researchers from this field tend to speak about \ZT{nanorobots} or \ZT{microrobots} rather than about \ZT{active particles}. General reviews covering medical applications of active particles (in the form of small robots) are given by Refs.\ \cite{OuLJWLWWTP2020,MedinaS2017,GiriMZ2021,Liu2019,CeylanYKHS2019,SittiCHGTYD2015,GaoWYLMH2021,WangTCP2020,SotoC2018,Wu2020,MujtabaEtAl2021,BuneaT2020,FuYZMKW2022,CeylanGKS2017,GuixMMM2014,WangKNZ2021,WuZCWXY2022,WangG2012,LiPYMMWHT2022,LiEFdAGZW2017,PengTW2017,SunWM2021}. Many reviews on medical applications of active particles and micromotors also focus on specific variants, such as acoustic nanodrops \cite{BordenSUS2020}, camouflaged motors \cite{GaoLLH2018}, catalytic motors \cite{SafdarKJ2018}, chemically propelled particles \cite{SomasundarS2021}, hydrogel-based systems \cite{WangRYW2021,LinXZLSM2020}, magnetic robots \cite{YangYCCZ2023}, motors with taxis behavior \cite{GaoFWTP2022}, and magnetic-resonance-imaging-guided particles \cite{VartholomeosFFM2011}.

A central medical application of active particles is \textit{drug delivery}. Standard drug delivery techniques face several challenges, such as the fact that the drugs also arrive in parts of the body where they are not supposed to arrive and cause damage there, or the fact that the targets of the drugs may be difficult to reach due to blood flow or biological barriers \cite{GaoW2014b}. In active drug delivery, drugs are therefore transported using active particles. A broad range of particle types with different propulsion mechanisms, such as chemical propulsion, light-driven propulsion, ultrasound-driven propulsion, propulsion by electric or magnetic fields, or even biological swimming (such as in sperm-driven biohybrid microrobots) can be used \cite{LuoFWG2018}. All these variants have their own advantages and disadvantages -- for instance, chemical propulsion mechanisms may rely on substances that are not biocompatible, light cannot penetrate deeply into biological tissue, microorganisms need nutrients, etc.\ \cite{LuoFWG2018}. Applications of active drug delivery include, among other things, the targeted transport of anticancer drugs, the transport of genes (in the form of nucleic acids) with the aim to treat inherited diseases or cancer, or the transport of insulin via glucose-responsive nanovehicles \cite{TezelTKGUOE2021}. Due to the importance of its field and the promises it has for practical applications, drug delivery via nanorobots is discussed in a large number of reviews \cite{LuYZQ2021,MedinaXS2018,ErkocYCYAS2019,TezelTKGUOE2021,GaoW2014b,KimGLF2018,PatraSDZPS2013,XuHGLQW2020,SrivastavaCAB2019,GhoshXGG2020,HuGCMQY2020,LuoFWG2018,XuWLYSM2020,WangLWPT2019,YoonYKGKMLC2020}.

Closely related is the topic of \textit{cancer treatment} based on active particles. An example for this is intravesical drug delivery in bladder cancer \cite{YoonYKGKMLC2020}. In cancer treatment, the aforementioned challenges of conventional medical approaches are particularly relevant -- cancer cells can be in places that are hard to reach, and conventional chemotherapy involves a number of severe adverse drug reactions due to its untargeted nature  \cite{SchmidtMES2020}. It is therefore advantageous to transport chemotherapeutic drugs directly to the cancer cells \cite{WangZ2021}. The treatment of cancer based on nanorobots is covered in a number of reviews \cite{YoonYKGKMLC2020,WangZ2021,SinghAM2023,SchmidtMES2020,SonntagSM2019,LiWM2020,GaoEFdAZW2018}.

Another application of medical nanorobots is \textit{diagnostics} and \textit{sensing}. For instance, one can attach biological receptors to the particles in order to detect the presence of biomolecules \cite{ChalupniakEtAl2015}. Moreover, the motion of the particles can also be used for diagnostic purposes -- for instance, the velocity of a chemically driven particle may depend on the concentration of certain chemicals and can therefore be used to determine this concentration \cite{KongDP2018,ZhengHZ2023}. In addition, it has been proposed to use active particles as medical imaging agents \cite{vanMoolenbroekPLS2020}.

Medical applications of active particles do not stop there. For instance, microrobots could perform simple surgery tasks, either autonomously or directed by a clinician \cite{NelsonKA2010}. Moreover, they can remove materials or act as controllable structures by opening or closing passageways \cite{NelsonKA2010}, for example by opening cell membranes \cite{WangWH2020}. They can also be applied in regenerative medicine, where they are used for cell delivery or induced cell proliferation \cite{LiuGP2022}. Moreover, microrobots have a variety of potential applications in the treatment of cardiovascular diseases \cite{GunduzAO2021}.

\section{Conclusions}
In this article, we have provided an overview over the field of active matter by discussing the existing review articles on this topic. Besides general reviews, there exists a great number of review articles focusing on specific subtopics such as active colloids, theoretical methods, collective dynamics, microswimmers, biological active matter, molecular active matter, artificial active matter, and medical applications. The existence of several hundred reviews published in just a few years demonstrates the importance that this field of research has gained and the pace at which it is still growing. 

\acknowledgments{M.t.V.\ and R.W.\ are funded by the Deutsche Forschungsgemeinschaft (DFG, German Research Foundation) -- Project-IDs 525063330 and 433682494 -- SFB 1459.}

\bibliography{refs,control}
\end{document}